\documentclass[jap,aip,showpacs,twocolumn]{revtex4}
\usepackage{amssymb}
\usepackage{amsmath}
\usepackage{epsfig}
\usepackage{graphicx}

\begin{document}

\title{First-principles study of spin-transfer torque in Co$_{2}$MnSi/Al/Co$_{2}$MnSi spin-valve}
\author{Ling Tang}\email{lingtang@zjut.edu.cn}
\author{Zejin Yang}
\affiliation{Department of Applied Physics, Zhejiang University of
Technology, Hangzhou 310023, P.~R.~China}

\begin{abstract}
The spin-transfer torque (STT) in Co$_{2}$MnSi(CMS)/Al/Co$_{2}$MnSi
spin-valve system with and without interfacial disorder is studied
by a first-principles noncollinear wave-function-matching method. It
is shown that in the case of clean interface the angular dependence
of STT for CoCo/Al (the asymmetry parameter $\Lambda\approx4.5$) is
more skewed than that for MnSi/Al ($\Lambda\approx2.9$), which
suggests the clean CoCo/Al architecture is much more efficient for
the application on radio frequency oscillation. We also find that
even with interfacial disorder the spin-valve of half-metallic CMS
still has a relatively large parameter $\Lambda$ compared to that of
conventional ferromagnet. In addition, for clean interface the
in-plane torkance of MnSi/Al is about twice as large as that of
CoCo/Al. However, as long as the degree of interfacial disorder is
sufficiently large, the CoCo/Al and MnSi/Al will show approximately
the same magnitude of in-plane torkance. Furthermore, our results
demonstrate that CMS/Al/CMS system has very high efficiency of STT
to switch the magnetic layer of spin-valve.
\end{abstract}


\maketitle

\section{Introduction}

The spin-transfer torque (STT) effect\cite{Slonczewski96,Berger96}
has attracted considerable interest in the recent years due to its
potential applications on future spintronics devices, such as the
STT-based magnetic random access memory
(STT-MRAM)\cite{MRAMS1,MRAMS2} and spin-transfer nanocontact
oscillator (STNO).\cite{STNO1,STNO2} However, in STT-MRAM the
biggest challenge is to reduce the switching current density that
mainly depends on the efficiency of current-induced STT, i.e. the
STT per current. Writing information by STT in MRAM with low
switching current can save power consumption and shrink the size of
memory cell, which are of crucial importance to realistic
applications. Another important issue is the angular dependence of
STT in STNO devices, where the asymmetry parameter $\Lambda$
dominates the output power of high-frequency
generation.\cite{Slonczewski96,Slonczewski02,Brataas06}

In order to reduce the critical current, the Co-based Heusler
alloys, such as Co$_{2}$MnSi (CMS) and Co$_{2}$FeAl (CFA) etc., are
believed as good candidates for STT-based memory
devices.\cite{bai13} These Heusler compounds not only have small
damping constant, low saturation
magnetization\cite{Mizukami09,Oogane10} and perpendicular magnetic
anisotropy,\cite{Li10,wang10,Li11} but also have been theoretically
predicted to be half-metallic materials with very high spin
polarization (P$\approx1$),\cite{deGroot83} which are all in favour
of decreasing critical current.\cite{Slonczewski96,sun00} Compared
to the early all-metallic spin-valve (SV) of conventional
ferromagnet such as Co/Cu multilayers system,\cite{Tsoi98,Myers99}
the magnetoresistance (MR) ratio and the efficiency of STT will be
enhanced for devices fabricated by half-metallic Heusler alloys due
to the high spin polarization. On the other hand, the magnetic
tunnel junction (MTJ) with Heusler electrodes also shows very large
tunneling MR ratio even at room temperature.\cite{TMR1,TMR2,TMR3}
However, the main disadvantage of MTJ devices is the large
resistance-area product (RA) due to the insulator spacer. So the
read-write speed of memory will be reduced for its large RC
constant. Meanwhile, the RA of all-metallic structure is apparently
much smaller, suggesting that the all-metallic SV of half-metallic
Heusler alloys has greater potential for spintronics applications.

Recently, the half-metallic SV system composed of CMS has received
increasing attention\cite{GMR1,GMR2,GMR3,GMR4} because the Curie
temperature of CMS is very high (985K) and the lattice of CMS can be
well matched with several nonmagnetic metals. The perpendicular
magnetic anisotropy of CMS also has been reported in CoPt/CMS hybrid
electrode,\cite{Hiratsuka10} which implies that it is suitable for
application on high-frequency generation.\cite{Okura11,Seki13} So
far, for CMS/Al/CMS trilayers there have been investigations that
focus on the conductance and its relationship with the interface
architectures.\cite{miura11,yu12} But all these studies do not
include the transport properties for disorder interface and the
first-principles calculation of STT in CMS/Al/CMS SV system has not
been reported up to now. Further, the dependence of STT on CMS/Al
interface architectures is still unclear yet. Therefore, the main
purpose of this paper is to study the magnitude and the
angle-dependence of STT in CMS/Al/CMS SV system by first-principles
method. In addition, the calculations will be carried out in
different CMS/Al interfaces, i.e. Co-terminated (CoCo/Al) and
MnSi-terminated (MnSi/Al) interface including alloyed interfacial
disorder.

\section{Computational Method}

In this paper, we employ the tight-binding linearized
muffin-tin-orbital (TB-LMTO) surface Green function method with
atomic sphere approximation (ASA) to obtain the effective
single-electron potential of CMS/Al(001)
interface.\cite{turek97,andersen85} The coherent potential
approximation (CPA)\cite{turek97} is introduced to deal with the
alloyed interfacial disorder. With the rigid potential
approximation,\cite{shuai08} the effective potential for atomic
spheres in fixed magnet of SV is rotated an angle $\theta$ relative
to the free magnet in spin space. So the STT can be calculated by
the corresponding noncollinear scattering wave-function with TB-LMTO
basis.\cite{xia06,shuai08,xu08,tang13}

In our calculation the CMS/Al/CMS/Al(001) SV system is composed of a
fixed semi-infinite CMS as left lead, a spacer layer with 9 Al
monolayers (MLs), a free ferromagnetic layer with 13 CMS MLs and a
semi-infinite Al as right lead. Here the three CMS/Al interfaces are
assumed to be the same architectures. The magnetization of free CMS
layer is align with $z$ axis. The direction of magnetization for
left fixed CMS lead is rotated at a relative angle $\theta$ in
$x$-$z$ plane (in-plane). So the transport direction is along the
$y$ axis (out-of-plane). The other details of interface model and
electronic structure calculations can be found in our previously
work.\cite{tang12} For evaluating the STT with clean interface, we
use $960\times960$ $k$-points in full two-dimensional (2D) Brillouin
zone (BZ) to obtain the convergent results. However, considering the
computational cost, the out-of-plane STT in the case of disorder
interface is not included in this work. In addition, for disorder
interface the convergence of in-plane STT can be obtained by a
$\textbf{k}_{\|}$ sampling equivalent to $50\times50$ $k$-points in
2D BZ.

\section{Results and Discussion}

\begin{figure}
  \includegraphics[width=8.6cm, bb=0 0 532 757]{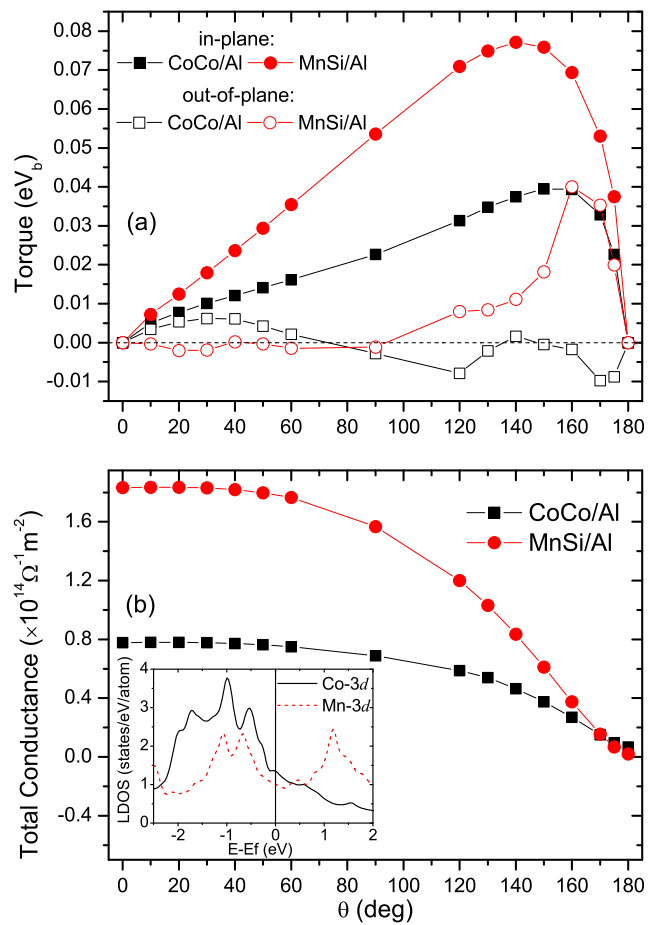}\\
  \caption{(color online). (a) The angular dependence of total STT on
   free CMS layer of CMS($\theta$)/Al/CMS/Al SV system
   with clean interface, where $\theta$ is relative angle between
   two magnetizations of CMS layer.
  Note that our calculated torque is generated by a small bias
  $V_{b}$ and just for one lateral unit cell with cross section
  area $A$.
 (b)
  The angular dependence of total conductance for the same structure.
  The inset shows the 3\emph{d} local density of states (LDOS) of interfacial Co atoms for CoCo/Al
  and Mn atoms for MnSi/Al. }\label{fig1}
\end{figure}

First, Fig.\ref{fig1} shows the calculated angular dependence of the
total STT on local spins in free CMS layer and the total conductance
in CMS/Al/CMS/Al SV system with clean CoCo/Al and MnSi/Al interface.
One can see that the in-plane STT deviates from the sine function
and the peak of in-plane STT is around $\theta=155^{\circ}$ for
CoCo/Al and $\theta=140^{\circ}$ for MnSi/Al. According to
Slonczewski model,\cite{Slonczewski02} these angular-dependent STT
can be fitted to the asymmetry parameter of $\Lambda\approx4.5$ for
CoCo/Al and $\Lambda\approx2.9$ for MnSi/Al. These values are much
larger than that of SV with conventional ferromagnet such as Co/Ni
multilayers system ($\Lambda\approx1.5$),\cite{Rippard10} which can
be attributed to the high spin polarization (P=1) originating from
the half-metallic nature of the CMS. Moreover, our calculated
results also suggest that the STNO device with clean CoCo/Al
interface will have much more output power of high-frequency
microwave compared to MnSi/Al.

\begin{figure}
  \includegraphics[width=8.6cm, bb=10 20 520 428]{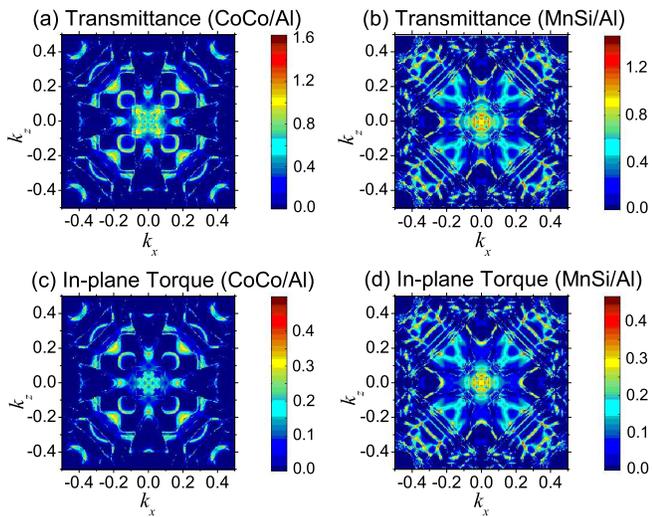}\\
  \caption{(color online). The in-plane wave vector $\mathbf{k}_{\parallel}$ dependence of
  transmittance (including spin-flip probability) and total in-plane STT on free CMS layer for
  CMS($\theta=150^{\circ}$)/Al/CMS/Al system with clean interface.
  }\label{fig2}
\end{figure}

\begin{figure}
  \includegraphics[width=8.6cm, bb=10 208 520 416]{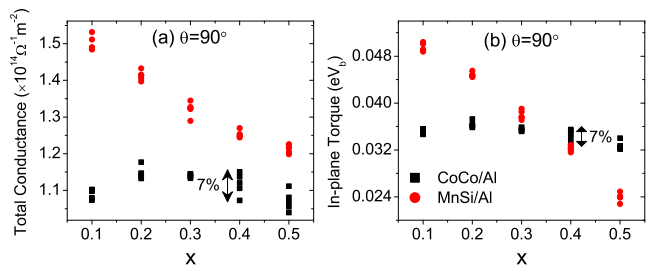}\\
  \caption{(color online). The total conductance and in-plane STT for different randomly
  generated configurations of disorder in the case of
  $\theta=90^{\circ}$, where $x$ is interfacial alloy concentration.
  It is shown that the maximum of uncertainties is about 7\% for both conductance
  and torque.
  }\label{fig3}
\end{figure}

The situation of out-of-plane STT for two kinds of interface is more
complicated. The out-of-plane STT of MnSi/Al has a peak at
$\theta=160^{\circ}$ and its magnitude is comparable with the
in-plane STT, which may have interesting consequences for the
switching process of the SV system. However, the out-of-plane STT of
CoCo/Al has no such peak structure as MnSi/Al and there are two sign
changes around $\theta=70^{\circ}$ and $140^{\circ}$, which is very
similar to the results of Co/Cu/Co
spin-valves.\cite{shuai08,haney07}

From Fig.\ref{fig1}(a), one can observe that the maximum of total
in-plane STT for MnSi/Al is about twice as large as that for
CoCo/Al. Note that our calculated torque is generated by a small
bias $V_{b}$ in the linear response regime.\cite{shuai08} Therefore,
the STT value shown in Fig.\ref{fig1}(a) denotes the zero bias
torkance that is defined as variation of the torque with the voltage
around zero bias. So for voltage biased applications of SV system,
Fig.\ref{fig1}(a) demonstrates the in-plane torkance for clean
MnSi/Al is much larger than that for clean CoCo/Al.

Similarly, Fig.\ref{fig1}(b) shows the obtained total conductance at
$\theta=0^{\circ}$ of MnSi/Al is also larger than that of CoCo/Al,
which is consistent with the previous first-principles
calculations.\cite{miura11,yu12} Miura \emph{et al.} have pointed
out\cite{miura11} that this conductance difference is due to the
larger 3\emph{d} component of LDOS at Fermi level for interfacial Co
atoms of CoCo/Al compared to that of interfacial Mn atoms of
MnSi/Al. Then the conduction electron will get stronger reflection
in CoCo/Al interface because the 3\emph{d}-orbital can be regard as
scatters of electron. Indeed, as shown in the inset of
Fig.\ref{fig1}(b), this fact is confirmed by our calculated results,
which also agrees with Yu \emph{et al.}'s calculation in CMS/Al
interface.\cite{yu12} In particular, the ratio of conductance
between MnSi/Al and CoCo/Al at $\theta=0^{\circ}$ is about 2.4,
which is very close to the ratio of maximum STT. Just as expected,
our calculations results show that the magnitude of STT is dependent
on interface transparency. Clearly, this dependence originates from
the fact that the larger electronic current will carry the more spin
moments to be absorbed by the local spins.

To reveal the interface transparency and STT in more detail, the
$\mathbf{k}_{\parallel}$ dependence of transmittance and the total
in-plane STT on free CMS layer for clean CoCo/Al and MnSi/Al
interface with $\theta=150^{\circ}$ are shown in Fig.\ref{fig2}.
Here the transmittance of each $\mathbf{k}_{\parallel}$ point
includes spin-flip probability. For both CoCo/Al and MnSi/Al
interface the transmittance and the total in-plane STT distribute
around the region of projected Fermi surface of CMS and Al. As
expected, the distribution of transmittance and the total in-plane
STT in 2D BZ is very similar to each other, which indicates that the
interface transparency dominates the magnitude of in-plane STT in
CMS/Al/CMS/Al SV system. In addition, from Fig.\ref{fig2} one can
observe that the states around $\Gamma$ point are really important
to the transport properties as discussions in previous
works,\cite{miura11,yu12} especially in MnSi/Al interface.

\begin{figure*}
  \includegraphics[width=16.5cm, bb=18 28 530 406]{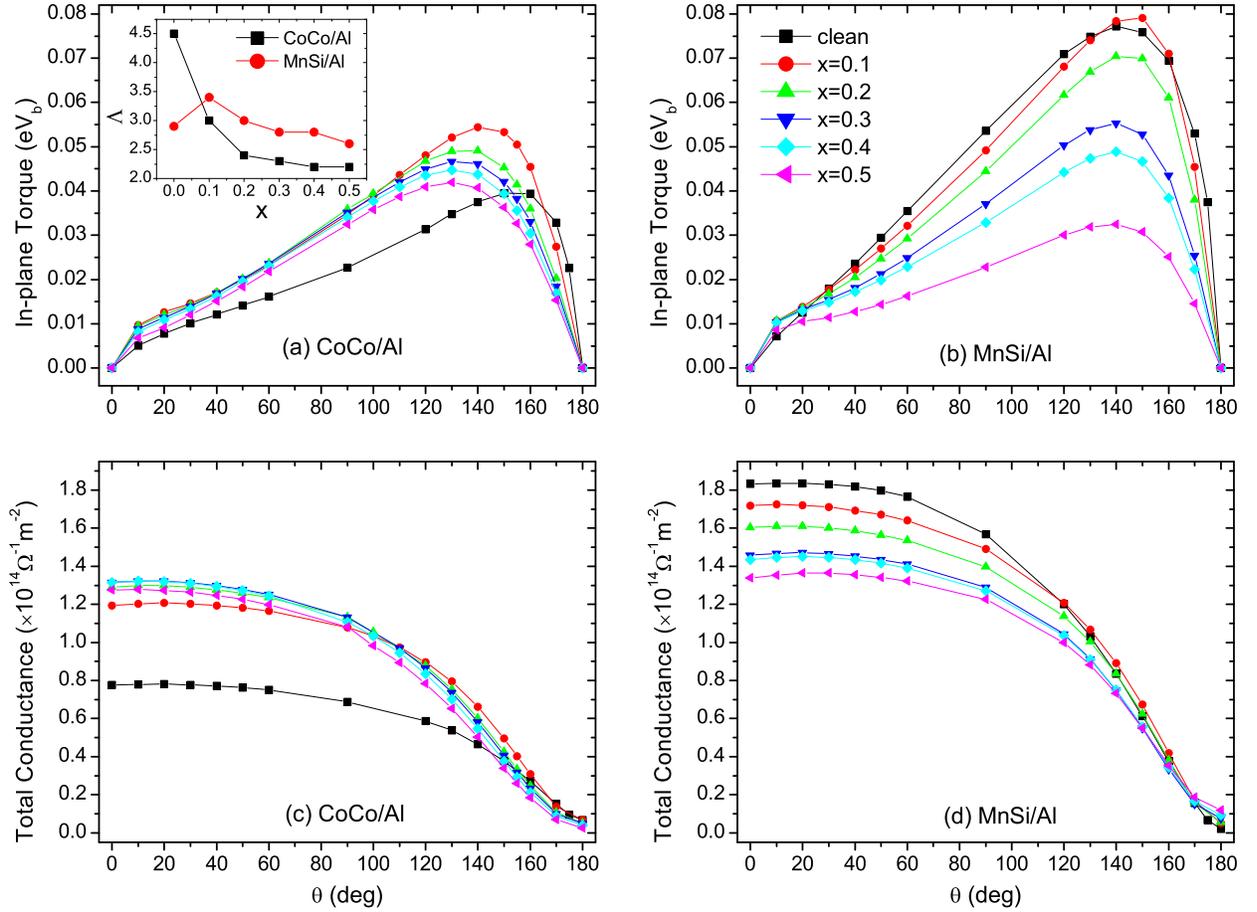}\\
  \caption{(color online). The angular dependence of total in-plane STT on free CMS
  layer
  and total conductance of
   CMS/Al/CMS/Al SV system with different interfacial alloy
   concentration $x$.
   It is shown that for large interfacial alloy concentration $x$
    the torkance (STT induced by unit bias) for
   MnSi/Al interface is approximately equal to that for
   CoCo/Al interface. The inset in figure (a) shows the
   concentration $x$ dependence of the estimated asymmetry parameter $\Lambda$
   of STT.
   }\label{fig4}
\end{figure*}

In realistic SV devices it is highly possible that the disorder will
exist at the interface. So in order to investigate the effect of
interfacial disorder on STT, we employ the model of substitutional
disorder alloy interface as in Ref.34. The details for the model of
disorder CMS/Al interface can also be found in Ref.37. Due to the
randomness of disorder configuration, first we have tested the
uncertainties of total conductance and total in-plane STT using five
CMS($\theta=90^{\circ}$)/Al/CMS/Al structures with randomly
generated disorder interface. Here the configurations of three
CMS/Al interfaces are also different. As shown in the
Fig.\ref{fig3}, the maximum spread is about 7\% for both conductance
and STT, which is close to the case of Co/Cu/Co system\cite{shuai08}
and have not substantially influence on the results about
concentration $x$ dependence.

Fig.\ref{fig4}(a) and (b) show the angular dependence of total
in-plane STT on free CMS layer of SV system with different
interfacial alloy concentration $x$. It can be see that the shape of
the angular dependence of STT becomes more symmetric with increasing
concentration $x$, which suggests that the output power of microwave
in CMS/Al system will be reduced by interfacial disorder. Here the
estimated parameter $\Lambda$ for different $x$ is shown in the
inset of Fig.\ref{fig4}(a). For CoCo/Al interface $\Lambda$
decreases rapidly as increasing concentration $x$, while for MnSi/Al
the influence of interfacial disorder on asymmetry parameters is
relatively small. However, even for interfacial disorder at
concentration $x=0.5$, the values of asymmetry parameters for both
interface are still relatively large ($\Lambda\approx2.2$ for
CoCo/Al and $\Lambda\approx2.6$ for MnSi/Al) compared to the case of
conventional ferromagnet.\cite{jia-prb11} Therefore, our results
indicate that the CMS/Al/CMS/Al SV system might be a good candidate
for STNO applications.

\begin{figure}
  \includegraphics[width=8.6cm, bb=18 0 541 755]{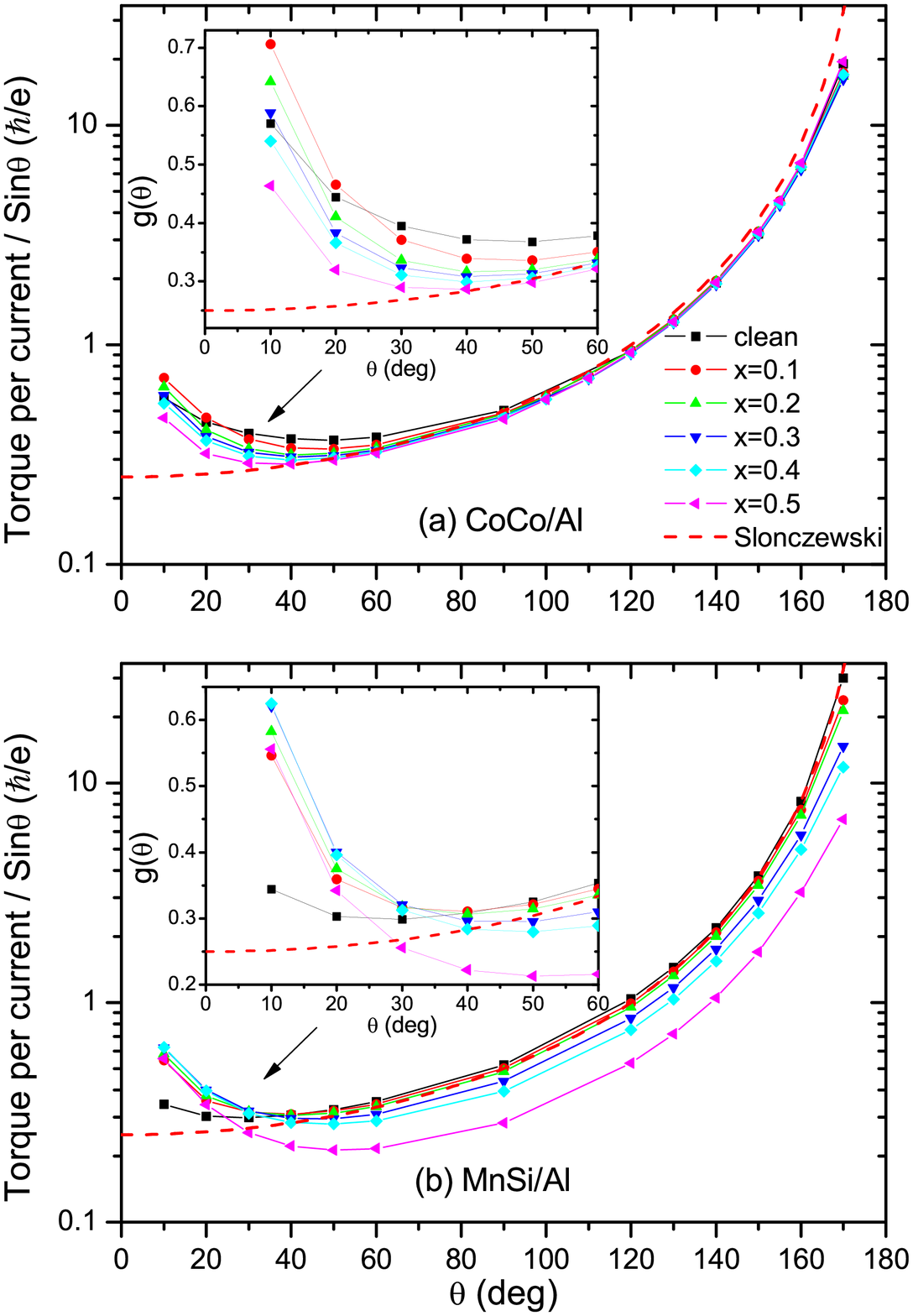}\\
  \caption{(color online). The angular dependence of efficiency parameter $g(\theta)$
   of STT, which is defined as the
  torque per current divided by $\sin\theta$.
  Note that at small angle the STT efficiency $g(\theta)$ for both CoCo/Al and MnSi/Al
  interface is much larger than the
  prediction of Slonczewski model (the red dash line) as shown
  in the inset.
 }\label{fig5}
\end{figure}

From Fig.\ref{fig4}(b) and (d), one can observe that the maximum of
STT and total conductance of parallel configuration for MnSi/Al
decrease monotonically with increasing $x$, except the in-plane STT
for $x=0.1$ is approximately equal to that for the clean interface.
However, contrary to the case of MnSi/Al, Fig.\ref{fig4}(a) shows
for CoCo/Al interface the maximum of STT will increase slightly once
the interfacial disorder has been introduced. Then for $x=0.5$ it
will decrease to the same value as that of clean interface.
Meanwhile, Fig.\ref{fig4}(c) shows the total conductance of parallel
configuration for CoCo/Al is significantly enhanced by the
interfacial disorder. A similar phenomenon is also observed in Fe/Cr
interface, where the conductance can be enhanced by a factor of
three.\cite{xia-prb01} This enhancement of total conductance can be
attribute to the electronic structure mismatch between the CMS and
Al layer.\cite{xia06} As the above discussion of clean interface,
the larger 3\emph{d} LDOS of interfacial Co atoms of CoCo/Al is
responsible for the higher reflection of propagating electrons. With
the mixture of Al atoms in the interfacial region, the reflection of
incoming electron will be reduced due to the 3\emph{d} orbital of Al
atoms is empty. Interestingly, despite the SV with clean MnSi/Al
will obtain larger torkance and conductance compared to clean
CoCo/Al, our calculated results show that when the degree of
interfacial disorder is sufficiently large, these two interfaces
will have approximately the same magnitude of torkance and
conductance.

\begin{figure}
  \includegraphics[width=8.6cm, bb=4 18 529 745]{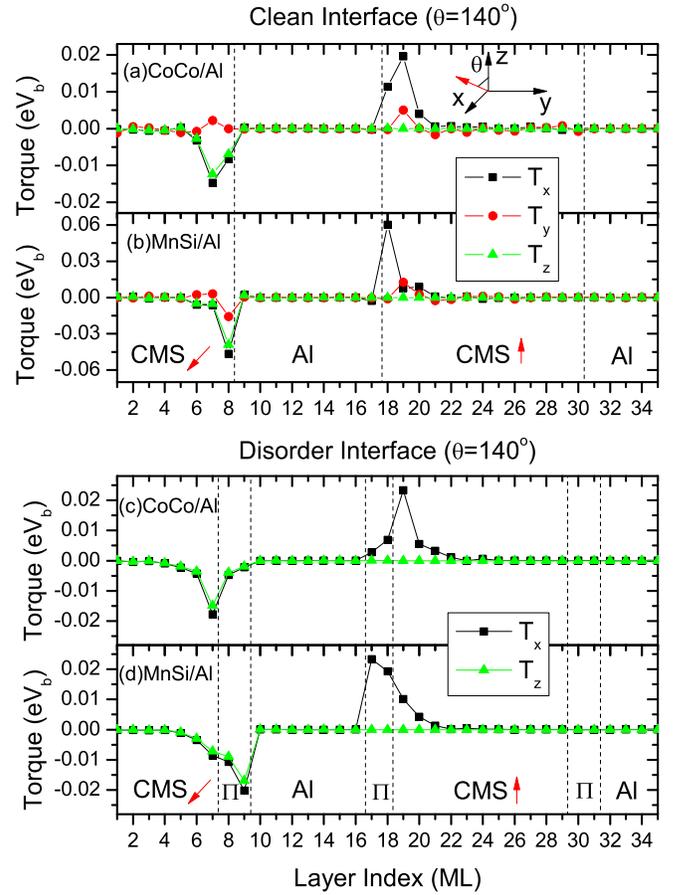}\\
  \caption{(color online). The layer dependence of STT on interfacial unit cell of
  CMS/Al/CMS/Al SV system
with clean and disorder interface, where $\Pi$ denotes the
interfacial layers with substitutional alloy disorder at $x=0.5$.
Here the red arrows indicate the
 magnetization
 of two CMS layers with relative angle $\theta$.
}\label{fig6}
\end{figure}

For the current-induced switching of magnetization in SV system, it
is believed that the torque per current is more important parameter.
Note that the above calculated result of STT is just in-plane
torkance $T_{in}(\theta)$ per lateral unit cell. Considering that
the area of cross section is $A$, the torque per current can be
written as $T_{in}(\theta)/[A\times G(\theta)]$, where $G(\theta)$
is the total conductance. So the parameter $g(\theta)$ for
describing STT efficiency can be defined as $g(\theta)\equiv
T_{in}(\theta)/[A\times G(\theta)]/\sin\theta$, where $A\times
G(\theta)$ is electronic current $I$ that flows through one lateral
unit cell at unit bias. This well-known parameter $g(\theta)$ is
just the deviation between the curve of angular dependence of STT
per current and the sine function, which represents the efficiency
of STT for switching the magnetic layer in spin-valve.

Here the calculated $g(\theta)$ for CoCo/Al and MnSi/Al interface
are shown in Fig.\ref{fig5}. As a comparison, the Slonczewski's
result\cite{Slonczewski96} with spin polarization P=1 is also
plotted in Fig.\ref{fig5}. For CoCo/Al with $\theta>60^{\circ}$ one
can observe that our obtained $g(\theta)$ is barely affected by
interfacial disorder and agrees well with the Slonczewski's result.
When $\theta$ is near $180^{\circ}$, our calculated $g(\theta)$ will
divergence as the prediction of Slonczewski, except the value is
smaller than Slonczewski's result. Similarly, the $g(\theta)$ of
MnSi/Al interface ($x\leq0.2$) is also well consistent with
Slonczewski's result for $\theta>60^{\circ}$. However, in contrast
to the case of CoCo/Al, the $g(\theta)$ of MnSi/Al ($x\geq0.3$) for
$\theta>60^{\circ}$ is significantly suppressed by interfacial
disorder, especially for the case of $x=0.5$.

As is well known, the inverses of $g(0^{\circ})$ and
$g(180^{\circ})$ are proportional to the critical current of
switching out of the parallel and anti-parallel states for SV. From
the inset of Fig.\ref{fig5}, one can see that the calculated
$g(\theta)$ close to $\theta=0^{\circ}$ are much larger than the
prediction of Slonczewski for both CoCo/Al and MnSi/Al.
Interestingly, in the case of MnSi/Al the $g(\theta)$ at small angle
is even larger for disorder interface compared to the clean
interface. In addition, one can also see that at $x=0.5$ the
$g(\theta)$ near $\theta=0^{\circ}$ has still relatively large value
for CoCo/Al ($g(10^{\circ})\approx0.47$) and for MnSi/Al
($g(10^{\circ})\approx0.55$). For the CMS/Al SV system, our
calculations show that the maximum of $g(\theta)$ at small $\theta$
approaches to 0.7, which is about one order larger than the
theoretical value of Co/Cu/Co SV.\cite{haney07} Therefore, our
calculated results suggest that the critical current of switching
magnetization in SV can be significantly reduced using half-metal
CMS as ferromagnetic layers.

Finally, to look the STT in CMS/Al SV a little further, we
demonstrate the layer resolved STT with clean and disorder interface
in Fig.\ref{fig6}. As expected, from Fig.\ref{fig6} one can observe
that the STT only exist around the interface region for both CoCo/Al
and MnSi/Al. Moreover, it will decay very fast and almost disappear
about 3-4 MLs away from the interface region, which indicates the
spin angular momentum is totally absorbed by the interface in CMS/Al
SV system. This behavior of layer resolved of STT is very similar to
the case of Fe/MgO/Fe tunnel juction\cite{stiles08}, where the
tunneling process is dominated by the half-metallic $\Delta_{1}$
state at $\Gamma$ point in Fe. Actually, in our CMS/Al/CMS SV system
all the states which are responsible for transport have
half-metallic nature, i.e. there is only propagating majority state
in CMS while the minority state is evanescent. Therefore, as
Heiliger and Stiles argued in Ref.42, the rapid decay of STT away
from CMS/Al interface can be attributed to the interference between
the evanescent minority and propagating majority in half-metal CMS.

\section{Summary}

In summary, the STT of CMS/Al/CMS/Al(001) SV system with and without
interfacial disorder have been studied by the noncollinear
wave-function matching method based on TB-LMTO with ASA
approximation. The obtained angular dependence of in-plane STT
deviates from the symmetric sine function. For clean interface we
estimate the asymmetry parameter $\Lambda\approx4.5$ for CoCo/Al and
$\Lambda\approx2.9$ for MnSi/Al, which indicates that the
CMS/Al-based high-frequency generators with clean CoCo/Al interface
is much more efficient. Furthermore, with increasing the degree of
interfacial disorder, the estimated value of $\Lambda$ will decrease
to $\sim2.2$ for CoCo/Al and $\sim2.6$ for MnSi/Al.

In addition, we also find that for the clean interface the in-plane
torkance of MnSi/Al is larger than that of CoCo/Al. However, once
the degree of interfacial disorder is sufficiently large, the
conductance and in-plane torkance of MnSi/Al will be approximately
the same as that of CoCo/Al. The efficiency of STT defined as torque
per current divided by $\sin\theta$ is also calculated in this work.
We find that for all the clean and disorder interfaces the obtained
efficiency parameter $g(\theta)$ at small $\theta$ are larger than
the prediction of Slonczewski's theory, where the maximum approaches
to 0.7. It suggests that the half-metallic CMS is a very good
candidate for STT-MRAM applications.

{\bf Acknowledgements} The authors are grateful to Prof. Ke Xia, Dr.
Shuai Wang and Dr. Yuan Xu for their preliminary works. We are also
grateful to: Ilja Turek for his TB-LMTO-SGF layer code; Anton
Starikov for the TB-MTO code based upon sparse matrix techniques.
The authors are thankful for financial supports from the National
Natural Science Foundation of China (Grant No:11104247 and
11304279), China Postdoctoral Science Foundation (Grant
No:2012M520666).


\begin{thebibliography}{}


\bibitem{Slonczewski96}J. C. Slonczewski, J. Magn. Magn. Mater. {\bf 159},
L1 (1996).

\bibitem{Berger96}L. Berger, Phys. Rev. B {\bf 54}, 9353 (1996).


\bibitem{MRAMS1}T. Kawahara, R. Takemura, K. Miura, J. Hayakawa, S. Ikeda, Y. M.
Lee, R. Sasaki, Y. Goto, K. Ito, T. Meguro, F. Matsukura, H.
Takahashi, H. Matsuoka and H. Ohno, IEEE J. Solid-St. Circ. {\bf
43}, 109 (2008).

\bibitem{MRAMS2}R. Takemura, T. Kawahara, K. Miura, H. Yamamoto, J. Hayakawa, N.
Matsuzaki, K. Ono, M. Yamanouchi, K. Ito, H. Takahashi, S. Ikeda, H.
Hasegawa, H. Matsuoka and H. Ohno, IEEE J. Solid-St. Circ. {\bf 45},
869 (2010).





\bibitem{STNO1}A. Ruotolo, V. Cros, B. Georges, A. Dussaux, J. Grollier, C.
Deranlot, R. Guillemet, K. Bouzehouane, S. Fusil and A. Fert, Nat.
Nano. {\bf 4}, 528 (2009).

\bibitem{STNO2}A. Slavin, Nat. Nano. {\bf 4}, 479 (2009).



\bibitem{Slonczewski02}J. Slonczewski, J. Magn. Magn. Mater. {\bf 247}, 324 (2002).

\bibitem{Brataas06}A. Brataas, G. E. W. Bauer, and P. J. Kelly, Phys. Rep. {\bf 427},
157 (2006).



\bibitem{bai13}Zhaoqiang Bai, Lei Shen, Guchang Han and Yuanping
Feng, SPIN {\bf 2}, 1230006 (2013).





\bibitem{Mizukami09}S. Mizukami, D. Watanabe, M. Oogane, Y. Ando, Y. Miura, M. Shirai
and T. Miyazaki, J. Appl. Phys. {\bf 105}, 07D306 (2009).

\bibitem{Oogane10}M. Oogane, T. Kubota, Y. Kota, S. Mizukami, H. Naganuma, A.
Sakuma and Y. Ando, Appl. Phys. Lett. {\bf 96}, 252501 (2010).






\bibitem{Li10}X. Q. Li, X. G. Xu, D. L. Zhang, J. Miao, Q. Zhan, M. B. A.
Jalil, G. H. Yu and Y. Jiang, Appl. Phys. Lett. {\bf 96}, 142505
(2010).

\bibitem{wang10}W. H. Wang, H. Sukegawa and K. Inomata, Appl. Phys.
Express {\bf 3}, 093002 (2010).

\bibitem{Li11}X. Q. Li, S. Q. Yin, Y. P. Liu, D. L. Zhang, X. G. Xu, J. Miao and Y. Jiang, Appl. Phys.
Express {\bf 4}, 043006 (2011).





\bibitem{deGroot83}R. A. de Groot, F. M Mueller, P. G. van Engen, and K. H. J.
Buschow, Phys. Rev. Lett. {\bf 50}, 2024 (1983).



\bibitem{sun00}J. Z. Sun, Phys. Rev. B {\bf 62}, 570 (2000).


\bibitem{Tsoi98}M. Tsoi, A. G. M. Jansen, J. Bass, W.-C. Chiang, M. Seck, V. Tsoi
and P. Wyder, Phy. Rev. Lett. {\bf 80}, 4281 (1998).

\bibitem{Myers99}E. B. Myers, D. C. Ralph, J. A. Katine, R. N. Louie and R. A.
Buhrman, Science {\bf 285}, 867 (1999).






\bibitem{TMR1}T. Ishikawa, S. Hakamata, K.-i. Matsuda, T. Uemura and M. Yamamoto, J. Appl. Phys.
{\bf 103}, 07A919 (2008).

\bibitem{TMR2}S. Tsunegi, Y. Sakuraba, M. Oogane, K. Takanashi and Y. Ando, Appl. Phys. Lett. {\bf
93}, 112506 (2008).


\bibitem{TMR3}W. Wang, E. Liu, M. Kodzuka, H. Sukegawa,
M. Wojcik, E. Jedryka, G. H. Wu, K. Inomata, S. Mitani and K. Hono,
Phys. Rev. B {\bf 81}, 140402 (2010).




\bibitem{GMR1}K. Yakushiji, K. Saito, S. Mitani, K. Takanashi, Y. K. Takahashi,
and K. Hono, Appl. Phys. Lett. {\bf 88}, 222504 (2006).


\bibitem{GMR2}T. Mizuno, Y. Tsuchiya, T. Machita, S. Hara, D. Miyauchi, K.
Shimazawa, T. Chou, K. Noguchi, and K. Tagami, IEEE Trans. Magn.
{\bf 44}, 3584 (2008).

\bibitem{GMR3}T. Iwase, Y. Sakuraba, S. Bosu, K. Saito, S. Mitani, and K.
Takanashi, Appl. Phys. Express {\bf 2}, 063003 (2009).


\bibitem{GMR4}Y. Sakuraba, K. Izumi, T. Iwase, S. Bosu, K. Saito, K. Takanashi, Y. Miura,
K. Futatsukawa, K. Abe, and M. Shirai, Phys. Rev. B {\bf 82}, 094444
(2010).



\bibitem{Hiratsuka10}T. Hiratsuka, G. Kim, Y. Sakuraba, T. Kubota, K. Kodama, N. Inami,
H. Naganuma, M. Oogane, T. Nakamura, K. Takanashi and Y. Ando, J.
Appl. Phys. {\bf 107}, 09C714 (2010).








\bibitem{Okura11}R. Okura, Y. Sakuraba, T. Seki, K. Izumi, M. Mizuguchi and K.
Takanashi, Appl. Phys. Lett. {\bf 99}, 052510 (2011).


\bibitem{Seki13}Takeshi Seki, Yuya Sakuraba, Ryo Okura, and Koki Takanashi, J.
Appl. Phys. {\bf 113}, 033907 (2013).


\bibitem{miura11}Y. Miura, K. Futatsukawa, S. Nakajima, K. Abe,
and M. Shirai, Phys. Rev. B {\bf 84}, 134432 (2011).

\bibitem{yu12}H. L. Yu, H. B. Zhang, X. F. Jiang, Y. Zheng, and G. W. Yang,
Appl. Phys. Lett. {\bf 100}, 222407 (2012).






\bibitem{turek97}I. Turek, V. Drchal, J. Kudrnovsk M. b, and P. Weinberger,
{\it Electronic Structure of Disordered Alloys, Surfaces and
Interfaces} (Kluwer, Dordrecht, 1997).

\bibitem{andersen85}O. K. Andersen, O. Jepsen, and D. Gloel, in {\it Highlights of
Condensed Matter Theory}, edited by F. Bassani, F. Fumi, and M. P.
Tosi (North-Holland, Amsterdam, 1985), pp. 59-176.



\bibitem{shuai08}S. Wang, Y. Xu and K. Xia, Phys. Rev. B {\bf
77}, 184430 (2008).



\bibitem{xia06}K. Xia, M. Zwierzycki, M. Talanana, P. J. Kelly and G. E. W. Bauer,
Phys. Rev. B {\bf 73}, 064420 (2006).



\bibitem{xu08}Y. Xu, S. Wang and K. Xia, Phys. Rev. Lett. {\bf 100}, 226602 (2008).


\bibitem{tang13}L. Tang, Z. J. Xu and Z. J. Yang, Int. J. Mod. Phys.
B {\bf 27}, 1350092 (2013).


\bibitem{tang12}L. Tang, Mod. Phy. Lett. B {\bf 26}, 1250205 (2012).



\bibitem{Rippard10}W. H. Rippard, A. M. Deac, M. R. Pufall, J. M. Shaw, M. W.
Keller, S. E. Russek, G. E. W. Bauer, and C. Serpico, Phys. Rev. B
{\bf 81}, 014426 (2010).



\bibitem{haney07}P. M. Haney, D. Waldron, R. A. Duine, A. S. N\'{u}$\tilde{\mathrm{n}}$ez, H. Guo, and A.
H. MacDonald, Phys. Rev. B {\bf 76}, 024404 (2007).



\bibitem{jia-prb11}Xingtao Jia, Ying Li, Ke Xia and Gerrit E. W. Bauer, Phys. Rev. B {\bf 84},
134403 (2011).



\bibitem{xia-prb01}K. Xia, P. J. Kelly, G. E. W. Bauer, I. Turek, J. Kudrnovsk\'{y} and
V. Drchal, Phys. Rev. B {\bf 63}, 064407 (2001).



\bibitem{stiles08}Christian Heiliger and M. D. Stiles, Phys. Rev. Lett. {\bf 100},
186805 (2008).







\end{thebibliography}
\end{document}